\documentclass[12pt,draft,onecolumn]{IEEEtran}
\usepackage{amsmath}

\title{Finding roots of polynomials over finite fields}
\author{Sergei V. Fedorenko, Peter V. Trifonov}
\date{July 24, 2001}

\begin{document}
\newtheorem{thm}{Statement}
\newtheorem{lem}{Lemma}
\newtheorem{defn}{Definition}
\newtheorem{sample}{Example}

\markboth{IEEE Transactions On Communications, Vol. 50, No. 11, NOVEMBER
2002}
{Fedorenko and Trifonov: Finding roots of polynomials over finite 
fields} 

\maketitle
\begin{abstract}In this paper we propose an improved algorithm for
finding roots of polynomials over finite fields. This makes
possible significant speed up of the decoding process of BCH, Reed-Solomon and
some other error-correcting codes.
\end{abstract}
\newcommand{\binomial}[2]{\begin{pmatrix}
			 #1\\
			 #2\end{pmatrix}}

\section{Introduction}

It is well known that one of the most time-consuming stages of
decoding process of Reed-Solomon, BCH and some other codes is
finding roots of the error-locator polynomial.  The most widely
known root finding algorithm is Chien search method, which is a
simple substitution of all elements of the field into the
polynomial, so it has very high time complexity for the case of
large fields and polynomials of high degree.

In \cite{HybridSolver} it was shown that every polynomial of
degree not higher than 5 can be transformed into a canonical form
with one or two parameters, so it is possible to construct tables
for finding roots. Moreover, if some roots are located in the same
cyclotomic coset, it is possible to eliminate them using
Euclidean algorithm.  In their recent paper \cite{Truong2001fast}
Truong, Jeng and Reed proposed a transformation which allows
grouping of some summands of the polynomial of degree not higher
than 11 into multiples of affine polynomials. Since affine
polynomials can be easily evaluated using very small pre-computed
tables, it is possible to speed up computations. However, their
algorithm suffers from some drawbacks:
\begin{enumerate}
\item It can be applied only to polynomials of degree not higher
than 11;
\item  Transformation of the polynomial is required.
Transformation proposed by authors ($y=x+f_6/f_7$ for polynomial
$F(x)=\sum_{i=0}^{11}f_ix^i$) can not be applied if $f_7=0$, so
root finding algorithm becomes more complicated;
\item After transformation the polynomial contains summand
$f_{10}y^{10}+f_9y^9$ (and $f_6x^6$ if transformation failed).
Evaluation of it still requires usage  of Chien's algorithm.
\end{enumerate}
In this paper we propose a common approach which can be used for
decomposition and fast evaluation of any polynomial.  We describe
it for the case of $GF(2^m)$, but our results can be generalized
for the case of arbitrary field. This technique can
be used in realization of Chien search.

Root finding problem can be formally stated as finding all distinct
$\displaystyle x_i: F(x_i)=0,\, F(x)=\sum_{j=0}^tf_jx^j, \\ x_i,f_j\in
GF(2^m)$. Chien search algorithm solves it by evaluation of $F(x)$ at all
$x\in GF(2^m)\backslash \mathbf{0}$ with the time complexity
\begin{equation}
\label{mChienComplexity}
W=(C_{add}+C_{mul})t(2^m-1),
\end{equation}
where $C_{add}$ and $C_{mul}$ are the time complexities of one addition
and multiplication in the finite field respectively. The algorithm
described below reduces cost of one polynomial evaluation using special
reordering of field elements.

\section{Fast polynomial evaluation algorithm} Before description
of the algorithm let us first consider some definitions and properties.
\begin{defn}
A polynomial $L(y)$ over $GF(2^m)$ is called a $p$-polynomial for
$p=2$ if
\[
L(y)=\sum_i L_iy^{2^i},\, L_i\in GF(2^m).
\]
\end{defn}
These polynomials are also called linearized polynomials.
The following
lemma describes the main property of $p$-polynomials.

\begin{lem}[\cite{Berlekamp1968Coding}]:
Let $y\in GF(2^m)$ and let $\alpha^0,\ldots,\alpha^{m-1}$ be a
standard basis. If \[y=\sum_{k=0}^{m-1}y_k\alpha^k, \,y_k\in
GF(2)\] and $L(y)=\displaystyle \sum_{j}L_jy^{2^j}$, then
\[L(y)=\sum_{k=0}^{m-1}y_kL(\alpha^k).\]
\end{lem}

A polynomial $A(y)$ over  $GF(2^m)$ is called  an affine polynomial if
$A(y)=L(y)+\beta,\,\beta\in GF(2^m)$, where $L(y)$ is a $p$-polynomial. The
above lemma makes possible evaluation of affine polynomials $A(x)$
with just one addition at each $x_i\in GF(2^m)$ if all $x_i$ are ordered
in their vector representation as Gray code.
\begin{defn}
Gray code is an ordering of all binary vectors of length $m$
such that only one bit changes from one  entry to the next.
\end{defn}
So if $x_i\in GF(2^m)$ are ordered as a Gray code \\ (i.e.
$wt(x_i-x_{i-1})=1$, where $wt(a)$ is the Hamming weight of $a$) the following holds:
\begin{equation*}
A(x_i)=A(x_{i-1})+L(\Delta_i),\, \Delta_i=x_i-x_{i-1}=\alpha^{\delta(x_i,x_{i-1})},
\end{equation*}
where $\delta(x_i,x_{i-1})$ indicates position in
which $x_i$ differs from $x_{i-1}$ in its vector representation. If $x_0=0$ then $A(x_0)=\beta$ and the
above equation describes the algorithm for evaluation of $A(x)$ at all points
of $GF(2^m)$.

\begin{sample}
Let us consider the case of $GF(2^3)$ defined by the primitive polynomial
$\pi(\alpha)=\alpha^3+\alpha+1$. One of many possible Gray codes is  the
sequence 000, 001, 011, 010, 110, 111, 101, 100 or
$0,1,\alpha^3,\alpha,\alpha^4,\alpha^5,\alpha^6,\alpha^2$. So one
needs to prepare a table of values $L(\alpha^0), L(\alpha^1),
L(\alpha^2)$. Then $A(1)=A(0)+L(\alpha^0)$,
$A(\alpha^3)=A(1)+L(\alpha^1)$ and so on.
\end{sample}

This algorithm can be applied for evaluation of any polynomial if it is
decomposed into a sum of affine multiples.

\begin{thm}
Each polynomial $F(x)=\sum_{j=0}^{t}f_jx^j, \\ f_j\in GF(2^m)$
can be represented as
\begin{equation*}
\label{mPolynomialDecomposition}
F(x)=f_3x^3+\sum_{i=0}^{\lceil(t-4)/5\rceil}x^{5i}(f_{5i}+\sum_{j=0}^{3}f_{
5i+2^j}\,x^{2^j}),
\end{equation*}
where $\lceil a\rceil$ is the smallest integer greater than or equal to
$a$.
\end{thm}
\begin{proof}
Let $k$ be the smallest integer such that $5k-1\geq t$ and assume that for
all $i>t \quad f_i=0$. Then the above equation  can be represented as
\begin{eqnarray*}
&F(x)=F_k(x)=f_3x^3+\\
&\sum_{i=0}^{k-2}x^{5i}(f_{5i}+\sum_{j=0}^{3}f_{
5i+2^j}x^{2^j})+\\&x^{5(k-1)}(f_{5(k-1)}+\sum_{j=0}^2f_{
5(k-1)+2^j}x^{2^j}).
\end{eqnarray*}
For $t=4$ ($k=1$)
this is obvious. Let us assume that $F_k(x)$ has been decomposed
as described. Then
$F_{k+1}(x)=F_k(x)+x^{5k}(f_{5k}+f_{5k+1}x+f_{5k+2}x^2+f_{5k+4}x^4)+x^{5(k-1)}f_{5(k-1)+8}x^8$.
The last summand of this expression can be
grouped with the last summand of the decomposition of $F_k(x)$.
\end{proof}
$p$-polynomials appearing in this decomposition have
only 4 summands. In some cases introducing additional summands can
reduce the total amount of affine polynomials in the final
decomposition.

So the whole root finding algorithm is as follows:
\begin{enumerate}
\item Compute $L_i^{(k)}=L_i(\alpha^k),\, k=[0;m-1], \\ i\in
[0;\lceil(t-4)/5\rceil]$, where $L_i(x)$ are $p$-polynomials appearing in the above decomposition: $L_i(x)=\sum_{j=0}^{3}f_{
5i+2^j}\,x^{2^j}$;
\item Initialize $A_i^{(0)}=f_{5i}$;
\item Represent each $x_j\in GF(2^m),\,j\in[0;2^m-1]$ in standard basis as
an element of Gray code with $x_0=0$, compute $A_i^{(j)}=A_i^{(j-1)}+L_i^{(\delta(x_j,x_{j-1}))},
j\in[1;2^m-1]$;
\item Compute $F(x_j)=f_3x_j^3+\sum_{i=0}^{\lceil
(t-4)/5\rceil}x_j^{5i}A_i^{(j)}, \\ j\in [1;2^m-1]$, and $F(0)=f_0$. If $F(x_j)=0$
then $x_j$ is a root of the polynomial. Note that the second summand of
this sum can be computed using Horner's rule.
\end{enumerate}

The total time complexity of this algorithm consists of complexity
of preliminary computations (first summand) and complexity of polynomial
evaluation and is equal to
\begin{eqnarray}
\label{mFastComplexity}
\nonumber
W_{fast}=&m\left\lceil\frac{t+1}{5}\right\rceil
(4C_{mul}+3C_{add})+ \\
&(\left\lceil\frac{t+1}{5}\right\rceil
(2C_{add}+C_{mul})+2C_{exp})(2^m-1),
\end{eqnarray}
where $C_{exp}$ denotes the time complexity of
one exponentiation over the finite field.

\section{Simulation results}
To demonstrate the efficiency of the  new algorithm it has been
implemented in C++ programming language, compiled with 
MS Visual C++ 6.0 
compiler  and software simulation on AMD 
Athlon 1700 XP  processor 
on Windows XP operating system
has been  performed. The multiplication of field elements in $GF(2^8)$ 
was implemented using  tables of logarithms and antilogarithms. The computation times required to evaluate the polynomials at the 
field elements $\alpha^0,\ldots,\alpha^{254}$ were averaged over 100000 computations and shown in Table 
1.

\begin{table*}[htb]
\label{tSimulations}
\caption{Computation time in microseconds for evaluating  the polynomials}
\begin{center}
\begin{tabular}{|c|c|c|c|c|}\hline
Degree&Chien search&TJR method&New method&New method speedup rate\\\hline
6&17.2&16.7&14.9&1.15\\\hline
7&19.8&18.2&15.1&1.31\\\hline
8&22.2&19.6&15.2&1.46\\\hline
9&24.6&20.3&15.3&1.60\\\hline
10&27.2&20.9&17.3&1.57\\\hline
11&29.6&20.6&18.2&1.62\\\hline
16&42.3&---&21.4&1.97\\\hline
24&61.8&---&25.8&2.39\\\hline
32&81.4&---&31.4&2.59\\\hline
\end{tabular}
\end{center}
\end{table*}

Note that speedup rates for Truong, Jeng and Reed method are
significantly lower than shown in  \cite{Truong2001fast}. This is caused
by different implementation of multiplication operation used in our simulations.

Comparing expressions \eqref{mChienComplexity} and \eqref{mFastComplexity}
and corresponding experimental results one can see that this algorithm can
be  up to 2.6 times faster  than Chien search depending on implementation of
operations over $GF(2^m)$.

\section{Conclusions}
In this paper we proposed an algorithm for evaluation of arbitrary polynomials at many points of the finite field with significantly
better performance than well-known Chien search. Sometimes
performance of this algorithm can be further improved by
construction of different polynomial decompositions.
\section*{Acknowledgements}
The authors would like to thank the transactions editor for
coding and communication theory application Prof. Vijay K. Bhargava
and the anonymous reviewers for their constructive comments.

The first author (S.~Fedorenko) would like to thank the
Alexander von Humboldt Foundation for support the work presented
in this paper.

\newpage
\bibliographystyle{IEEEbib}

\begin{thebibliography}{1}

\bibitem{HybridSolver}
R.T. Chien, B.D. Cunningham, and I.B. Oldham,
\newblock ``Hybrid methods for finding roots of a polynomial with application
to {BCH} decoding,''
\newblock {\em IEEE Transactions on Information Theory}, vol. 15, no. 2, pp.
329--335, 1969.

\bibitem{Truong2001fast}
T.-K. Truong, J.-H. Jeng, and I.S. Reed,
\newblock ``Fast algorithm for computing the roots of error locator polynomials
up to degree 11 in {Reed-Solomon} decoders,''
\newblock {\em IEEE Transactions on Communications}, vol. 49, no. 5, pp.
779--783, 2001.

\bibitem{Berlekamp1968Coding}
E.R. Berlekamp,
\newblock {\em Algebraic coding theory},
\newblock New York: McGraw-Hill, 1968.

\end{thebibliography}

\end{document}